\documentclass[aps,prl,twocolumn,groupedaddress,reprint,showpacs]{revtex4-2}
\usepackage{graphicx}
\usepackage{setspace}
\usepackage{amsmath}
\usepackage{amssymb}
\usepackage{color}
\usepackage{array}
\usepackage{hyperref}
\usepackage{float}
\usepackage{lipsum}
\usepackage{lineno}
\usepackage{subcaption} %

\usepackage[utf8]{inputenc}

\usepackage[all]{xy}
\newcommand{\RN}[1]{%
  \textup{\uppercase\expandafter{\romannumeral#1}}%
}

\begin{document}
\title{Predictive electronic feedback control for narrow-linewidth lasers}

\author{Jiatian Liu}
\thanks{Jiatian Liu and Xuan Qi contributed equally to this work.}
\affiliation{Shanghai Institute of Optics and Fine Mechanics, Chinese Academy of Sciences, Shanghai 201800, China}
\affiliation{University of Chinese Academy of Sciences, Beijing 100049, China}

\author{Xuan Qi}
\thanks{Jiatian Liu and Xuan Qi contributed equally to this work.}
\affiliation{Shanghai Institute of Optics and Fine Mechanics, Chinese Academy of Sciences, Shanghai 201800, China}
\affiliation{University of Chinese Academy of Sciences, Beijing 100049, China}

\author{Xiayang Fan}
\affiliation{Shanghai Institute of Optics and Fine Mechanics, Chinese Academy of Sciences, Shanghai 201800, China}
\affiliation{University of Chinese Academy of Sciences, Beijing 100049, China}

\author{Shenchao Jin}
\affiliation{Shanghai Institute of Optics and Fine Mechanics, Chinese Academy of Sciences, Shanghai 201800, China}
\affiliation{University of Chinese Academy of Sciences, Beijing 100049, China}

\author{Yuan Sun}
\email[email: ]{yuansun@siom.ac.cn}
\affiliation{Shanghai Institute of Optics and Fine Mechanics, Chinese Academy of Sciences, Shanghai 201800, China}
\affiliation{University of Chinese Academy of Sciences, Beijing 100049, China}

\begin{abstract}
Narrow-linewidth lasers are essential for various key atom-laser interaction processes and optical setups in quantum coherent control, quantum precision measurement, and quantum information science. The Pound-Drever-Hall (PDH) technique is the standard method for the frequency-stabilization of narrow-linewidth lasers, but the inherent time delay in its feedback loop inevitably induces excessive phase noise, commonly known as servo bumps. Previous solutions primarily rely on additional optical components outside the electronic feedback control loop. Here, we propose and implement a generic method based on the Smith predictor to address the time-delay issue entirely within the electronic part of the PDH feedback control loop. We systematically demonstrate its capability to suppress phase noise and reduce servo bumps by adding only a single Field-Programmable Gate Array (FPGA) controller to the loop. This class of methods is demonstrated to be conceptually concise with minimal hardware overhead, effective in practice, and holds potential for a wide variety of future applications.
\end{abstract}
\maketitle

A broad range of frontier research and mission-critical applications in quantum precision measurement and quantum information science rests fundamentally on high-performance laser sources \cite{PRXQuantum.6.010102}. Within this demanding landscape, narrow-linewidth performance is of particular importance as it is the decisive factor for various key atom-laser interaction processes and optical setups.
The relevant research fields include optical atomic clocks \cite{RevModPhys.87.637, PhysRevLett.123.033201, bothwell_jila_2019,  ohmae_transportable_2021, lin_87_2021}, optical nuclear clocks \cite{Th_nuclear_clock_nature_Ding, Th_nuclear_clock_nature_Schumm},  gravitational wave detection \cite{the_ligo_scientific_collaboration_advanced_2015, PhysRevLett.116.061102}, high-precision coherent synthesis of ultracold molecules \cite{takekoshi_ultracold_2014, molony_creation_2014, guo_creation_2016, Hexiaodong_science2020}, dark matter searches \cite{kobayashi_search_2022}, the detection of drifts in fundamental physical constants \cite{blatt_new_2008}, quantum interfaces \cite{casabone_heralded_2013, lee_ion-based_2019, covey_telecom-band_2019, covey_microwave--optical_2019, huie_multiplexed_2021}   and quantum simulation with atoms and molecules \cite{science_review2017, kokail_self-verifying_2019, joshi_quantum_2020, nphys_review2024}. 
In particular, Rydberg atoms have emerged as a promising physical platform in quantum sensing \cite{RevModPhys.89.035002, Holloway2024Review, Zhang2024Review} and quantum information science \cite{RevModPhys.82.2313, Saffman_review_2026} in which narrow-linewidth lasers play an essential role. In driving coherent ground-Rydberg transitions for quantum logic gates \cite{PhysRevA.92.022336, PhysRevA.105.042430, norrell_entangling_2026} or electric-field detection \cite{Jin2025CPL, Yuan_polarizatioin_wm}, suppression of laser phase noise \cite{PhysRevA.89.032516} across the spectrum from low-frequency to high-frequency is equally indispensable.

Narrow-linewidth lasers are typically realized by frequency-locking to ultra-stable optical reference cavity, and the Pound-Drever-Hall (PDH) technique has become the standard method for frequency discrimination and closed-loop control \cite{drever_laser_1983}. The PDH technique converts laser-frequency detuning from the reference-cavity resonance into a high-frequency electric error signal, enabling high-speed feedback correction of the laser frequency. From the perspective of the frequency-domain mechanism in control theory, the transmission delay $\tau$ of the signal in the optical path and electronic circuits introduces an additional phase lag that increases linearly with frequency $f$. Its magnitude can be expressed as: $\Delta \phi(f) = -2\pi f \tau$. When the noise frequency approaches the critical point of $1/(2\tau)$, the phase lag caused by this delay reaches $\pi$. At this point, the original negative feedback correction signal of the system undergoes a phase reversal, transforming into positive feedback that disrupts system stability. In other words, the feedback control bandwidth is practically constrained to the order of $1/\tau$. 
Constrained by this physical limit of finite bandwidth in the control loop, attempting to suppress low-frequency noise as much as possible to increase control gain will lead to a significant deterioration of the phase margin in the high-frequency band. The loop then not only fails to suppress noise near the bandwidth edge but actually amplifies its response to that noise. Consequently, this inevitably introduces a significant broadband noise amplification peak typically distributed in the range from tens of kHz to several MHz, commonly known as the servo bump.
The mid-frequency and high-frequency phase noises caused by the servo bumps exhibit distinctly different deleterious mechanisms across various quantum processes. 
For instance, in the single-qubit and two-qubit entangling gates of atomic qubits,  studies indicate that when the unsuppressed phase fluctuations extend to frequencies comparable to the Rabi frequency driving the qubit operations, they induce extra dephasing and state leakage \cite{PhysRevA.89.032516, PhysRevA.97.053803, PhysRevLett.121.123603, PhysRevA.107.042611, PhysRevApplied.20.L061002, norrell_entangling_2026}.

\begin{figure*}[t]
    \centering
    \begin{minipage}[t]{0.33\textwidth} 
        \centering
        \makebox[\linewidth][l]{\Large (a)}\vspace{-1pt}
        \includegraphics[width=\linewidth]{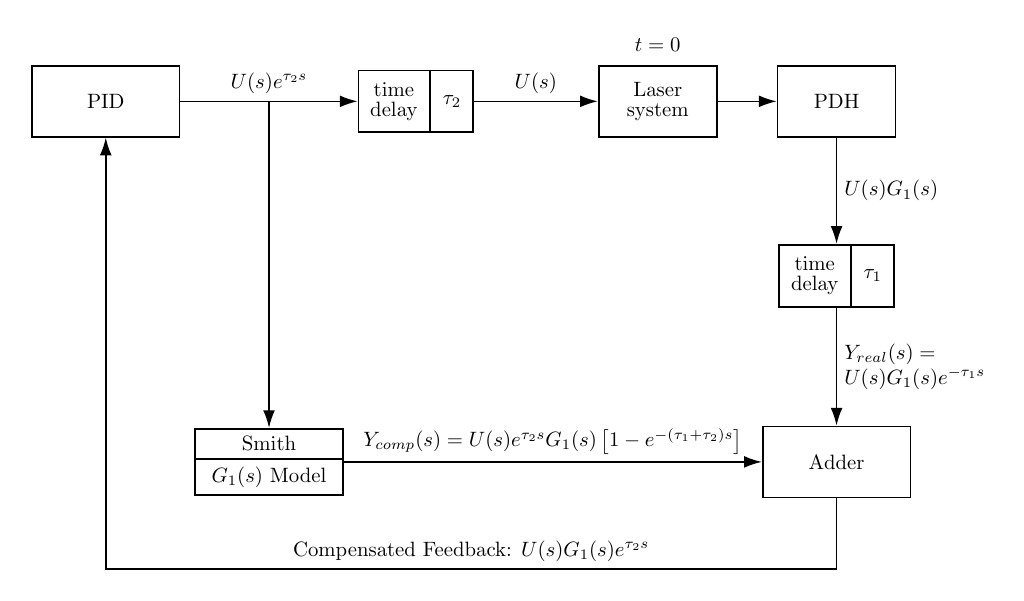}
        \label{fig:overall_sub1}
    \end{minipage}
    \hfill 
    \begin{minipage}[t]{0.37\textwidth}
        \centering
        \makebox[\linewidth][l]{\Large (b)}\vspace{-1pt}
        \includegraphics[width=\linewidth]{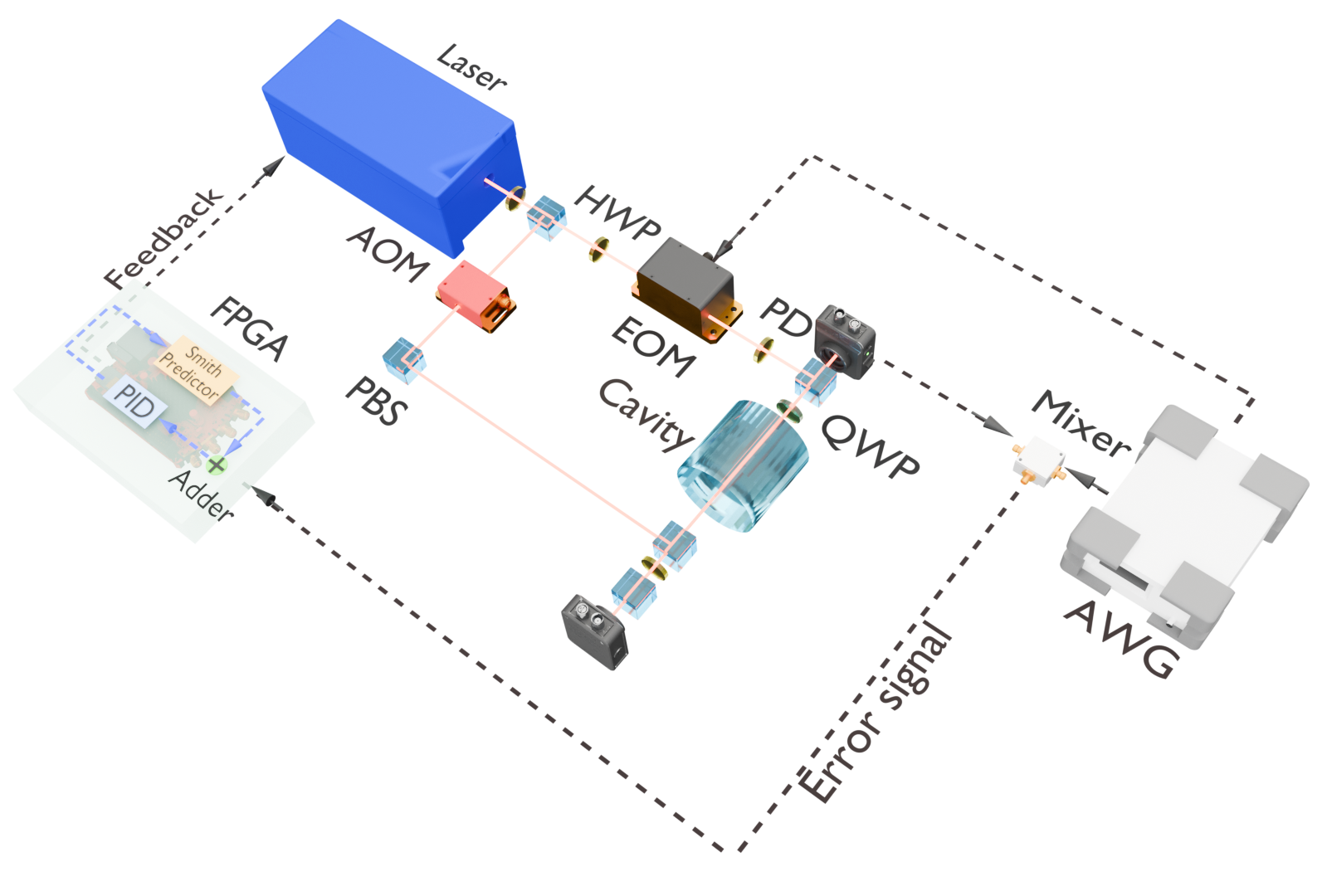}
        \label{fig:overall_sub2}
    \end{minipage}
    \hfill
    \begin{minipage}[t]{0.25\textwidth}
        \centering
        \makebox[\linewidth][l]{\Large (c)}\vspace{-1pt}
        \includegraphics[width=\linewidth]{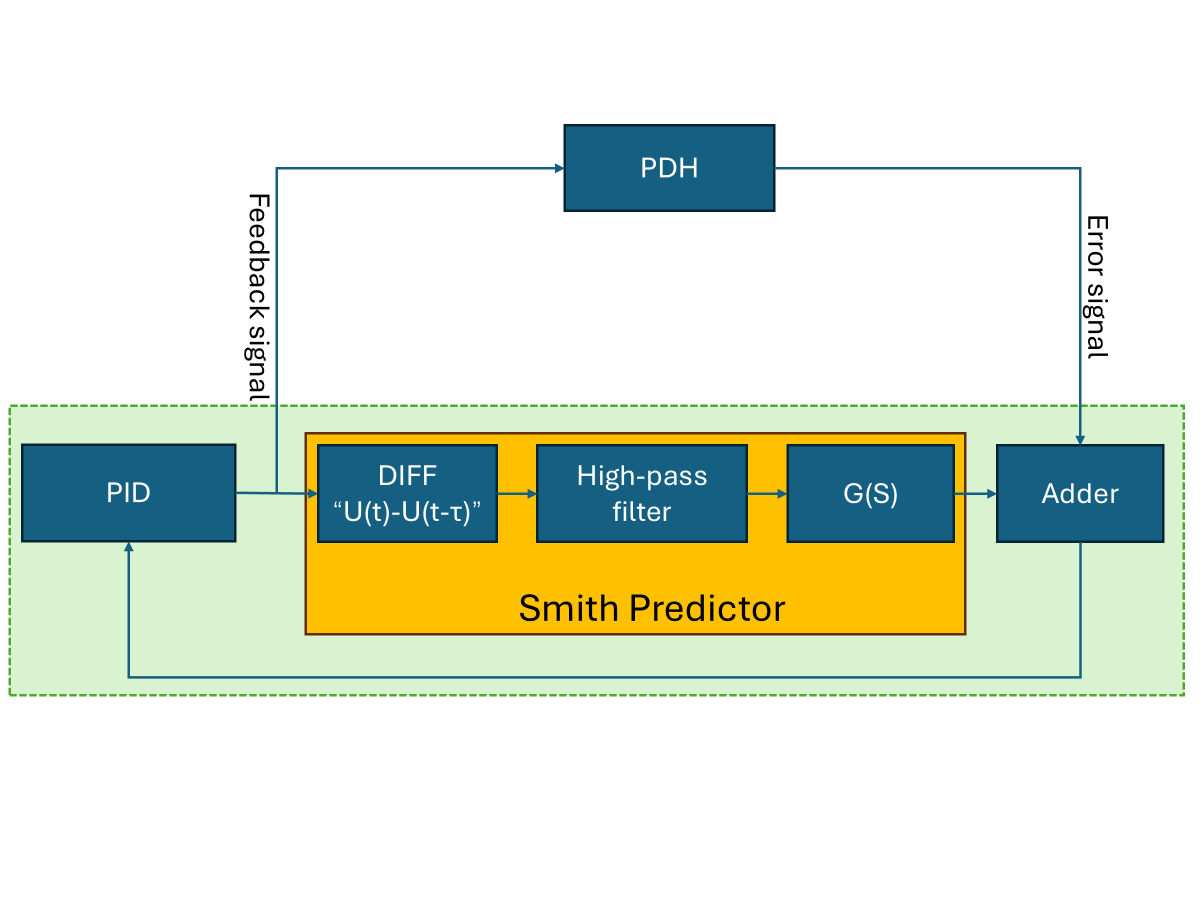}
        \label{fig:overall_sub3}
    \end{minipage}
    \caption{Basic design, experimental setup and control system architecture. (a) Theoretical block diagram of the closed-loop control system incorporating a Smith predictor, illustrating the inherent system time delays ($\tau_1$ and $\tau_2$) and the mathematical signal flow for delay compensation. $U(s)$ is the Laplace transform of the control signal $U(t)$ defined on the time domain $[0, \infty)$ for a prescribed starting time $t=0$. (b) Schematic of the experimental setup, demonstrating the integration of the Smith predictor for evaluating and suppressing system phase noise. HWP, Half-Wave Plate. QWP, Quarter-Wave Plate. PBS, Polarizing Beamsplitter. EOM, Electro-Optic Modulator. AOM, Acousto-Optic Modulator. AWG, Arbitrary Waveform Generator. PD, Photodiode. (c) Logical block diagram of the digital control loop. The functional blocks enclosed in the green box, including the Smith predictor and PID controller, are fully implemented within the FPGA architecture.}
    \label{fig:overall}
\end{figure*}

Previously, the suppression of excessive noises induced by the servo bump relies mainly on direct spectral filtering using optical cavities \cite{Hald_2005, akerman_simultaneous_2015, PhysRevA.97.053803, PhysRevLett.121.123603}, self-injection lock \cite{Krinner2024, IEEE10379078, Zhang:25}, or feedforward compensation techniques often with delay fibers \cite{PhysRevApplied.18.064005, YouLi2024Optica, PhysRevLett.133.253202, PhysRevA.111.042614, Nakamura2025}. 
However, these methods inevitably necessitate the introduction of additional optical components outside the electronic feedback loop, which usually limits the available output laser power or adds extra burden in tuning and stabilizing the system. 
Facing the future demands for narrow-linewidth lasers, we propose and implement a generic method to address the time-delay issue entirely within the electronic part of the PDH feedback control loop, which can be interpreted as a purely electronic suppression scheme for the servo bump. 
Without requiring any extra optical elements, this approach effectively reduces the amplitude of servo bumps solely by processing the error and feedback signals within the feedback loop. As the servo bump is an inherent characteristic of the operating control-feedback loop in a PDH lock, and although it is noise-like in nature, its properties remain invariant over time if the system is sufficiently stable and robust, predictive electronic feedback control then emerges as a viable solution. Specifically, by integrating the Smith predictor \cite{Smith1957CLOSERCO} from classical control theory with the PDH technique, and leveraging the real-time signal processing capabilities of an FPGA, our work successfully achieves effective suppression of the servo bump using a single controller.

We first present the theoretical analysis of delay compensation and the working principles of our method as shown in Fig. \ref{fig:overall}(a). 
The total loop delay comprises two distinct components $\tau_1, \tau_2$. $\tau_1$ represents the link delay starting from the control signal acting on the laser, propagating through the optical path and PDH demodulation, until the generated error signal is fed back before the Proportional-Integral-Derivative (PID) module. $\tau_2$ represents the forward transmission delay between PID and the laser system.
In the Laplace domain, let $s$ represent the complex frequency variable and let $U(s)$ denote the control signal received by the laser system. Due to the delay $\tau_2$, the signal output by the PID controller is mathematically equivalent to $U(s)e^{\tau_2 s}$. Assume that the laser frequency precisely follows the control of $U(s)$, then the output of PDH module becomes $U(s)G_1(s)$ with $G_1(s)$ denoting its ideal, delay-free transfer function. Subsequently, this signal undergoes delay $\tau_1$, and the physically generated error signal can be expressed as $Y_\text{real}(s) = U(s)G_1(s)e^{-\tau_1 s}$.
We introduce the Smith predictor into the loop and integrate it seamlessly with the laser frequency stabilization process. The PID output $U(s)e^{\tau_2 s}$ is split into two paths: the main path drives the laser system as usual, while the bypass is fed directly into the Smith predictor. Internally, the predictor performs a parallel calculation utilizing the known model of $G_1(s)$ and delivers a differential compensation signal $Y_\text{comp}(s) = U(s)e^{\tau_2 s}G_1(s)[1 - e^{-(\tau_1+\tau_2)s}]$.
An adder combines the physical signal influenced by delay $Y_\text{real}(s)$ with the compensation signal $Y_\text{comp}(s)$. Ideally, the output of adder $U(s)e^{\tau_2 s}G_1(s)e^{-(\tau_1+\tau_2)s}$ exactly equals to $U(s)G_1(s)e^{\tau_2 s}$, mathematically canceling out the actual physical signal $Y_\text{real}(s)$. Ultimately, the compensated feedback signal returned to the PID module is restored to $U(s)G_1(s)e^{\tau_2 s}$. 
Therefore, by updating the electronic and signal-processing parts of the entire system, the time delays are effectively compensated and therefore the servo bump would be suppressed.

The experimental setup of the proposed method is depicted in Fig. \ref{fig:overall}(b), and includes an optical beat note measurement to independently evaluate the system's phase noise. Specifically, the transmitted light from the ultra-stable optical cavity is optically beaten against the frequency-shifted first-order diffracted light generated by an acousto-optic modulator (AOM). The AOM is driven at a constant frequency of 80 MHz. Under these experimental conditions, the carrier power fluctuation of the resultant beat signal is strictly maintained below 1 dB, thereby establishing a stable baseline for subsequent noise characterization. We choose the closed-loop injection method utilizing a Frequency Response Analyzer (FRA) to obtain the frequency response characteristics of the PDH system with a high signal-to-noise ratio (SNR). Due to the system's extremely high open-loop gain in the low-frequency band, the effective measurement frequency band was set from 10 kHz to 5 MHz. Detailed experimental procedures regarding the perturbation signal injection and the dual-channel synchronous acquisition used to eliminate inherent equipment delays are provided in the Supplementary Material.

Building upon the characterized system response, the Smith predictor is digitally implemented to compensate for the pure time delay, the logical block diagram of which is illustrated in Fig. \ref{fig:overall}(c). The operations enclosed within the green-shaded region are executed within the internal architecture of the FPGA. Within this digital pipeline, the output from PID is split into two parallel pathways. The primary path carries the active feedback signal, which is routed through the physical PDH link to the laser's frequency control port to maintain frequency locking. Concurrently, the secondary path is directed into an internal ring buffer to compute the differential signal between the instantaneous control effort $U(t)$ and its delayed counterpart $U(t - \tau)$ with $\tau=\tau_1+\tau_2$.
Because the predictor model mismatch introduces a residual error proportional to $1 - e^{-\tau s}$, a second-order high-pass filter with a 20 kHz cutoff is inserted to bound this residual error at low frequencies.
After convolution with the plant model $G(s)$, the filtered predictive signal is superimposed onto the raw physical PDH error signal. This synthesis ultimately yields a delay-compensated error signal, which is then fed back to PID, thereby robustly closing the compensation loop.

In the digital logic implementation on FPGA, the aforementioned predicting process is accomplished based on an internal register stack. The system clock frequency is set to 125 MHz. The predictor buffers historical data through a shift register with a depth of $N=1024$ to simulate the pure time-delay element of the system, supporting a maximum delay compensation of approximately 8.2 $\mu\text{s}$. 
Assuming the computational time of the predictor model itself is negligible, the time-delay state can be reconstructed by extracting historical data at a specific depth. For instance, when the measured feedback link delay is 200 ns and the forward transmission delay is 208 ns, the corresponding numbers of clock cycles are 25 and 26, respectively. The system can then address and extract the register data at depth $n = 25 + 26 = 51$ to serve as the simulated delay variable $U(s)e^{\tau_2 s}G_1(s)e^{-(\tau_1+\tau_2)s}$. This value is utilized for subsequent differential and summation operations, thereby efficiently realizing the reconstruction of the delay-free error signal within the digital domain.  

\begin{figure}[h]
    \centering
    \includegraphics[width=\columnwidth]{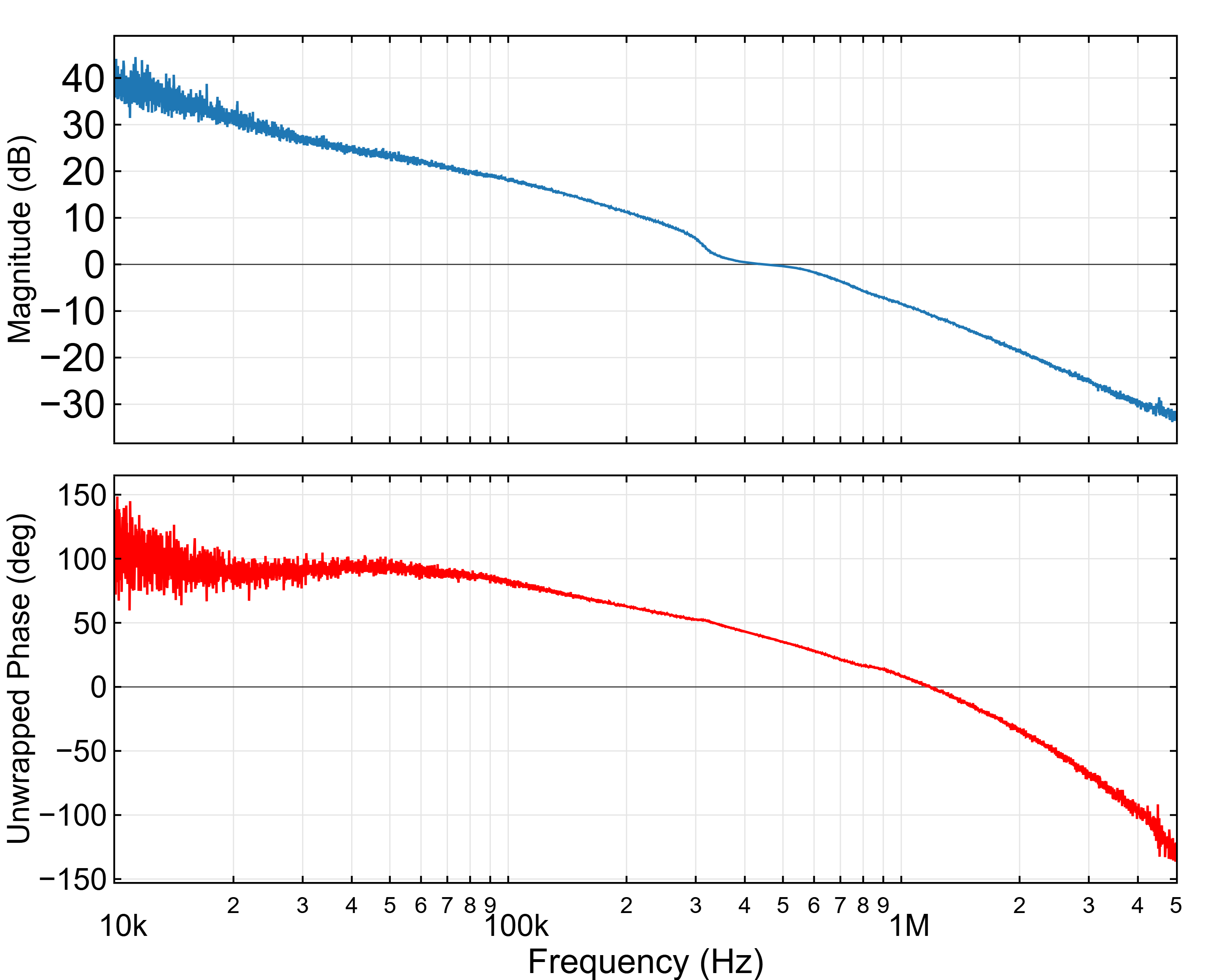}
    \caption{Top: Open-loop magnitude response of the plant, $\vert G(j\omega)\vert$. The blue line represents the plant gain (dB) extracted from the closed-loop system, illustrating its inherent frequency attenuation characteristics.
    Bottom: Open-loop phase response of the plant, $\angle G(j\omega)$. The red line denotes the unwrapped phase (deg), clearly demonstrating the linear phase lag induced by hardware time delay.}
    \label{fig:System Response}
\end{figure}

Here, the core of designing the Smith predictor is the accurate modeling of the dynamic response of the PDH feedback control loop including the laser. Therefore, the primary step requires characterizing the amplitude-frequency and phase-frequency responses. 
Experimentally, this test is performed by injecting a swept-frequency signal with known amplitude and frequency into the laser feedback terminal within the loop while simultaneously monitoring the error signal. By analyzing the transfer relationship between the error signal and the combination of the feedback and injected signals, the amplitude and phase response curves of the system can be derived.
A set of measured amplitude-frequency and phase-frequency response curves is shown in Fig. \ref{fig:System Response}. The PDH frequency stabilization loop exhibits a pronounced low-pass characteristic, with its amplitude attenuation behavior closely approximating a first-order integrating system. According to the phase-frequency response curve, the system phase stabilizes around $90^\circ$ in the low-frequency regime. Considering the inherent $180^\circ$ phase shift introduced by the negative feedback design and the asymptotic $-90^\circ$ phase lag characteristic of the open-loop system in the low-frequency regime, their combined effect yields the phase of $90^\circ$ at these frequencies. This aligns consistently with the high-gain characteristic observed at low frequencies in the amplitude-frequency curve.
In the higher frequency region, the time delay embedded in the system leads to a significant increase in phase lag that is linear with respect to frequency.

\begin{figure}[h!] 
    \centering
        \includegraphics[width=0.45\textwidth]{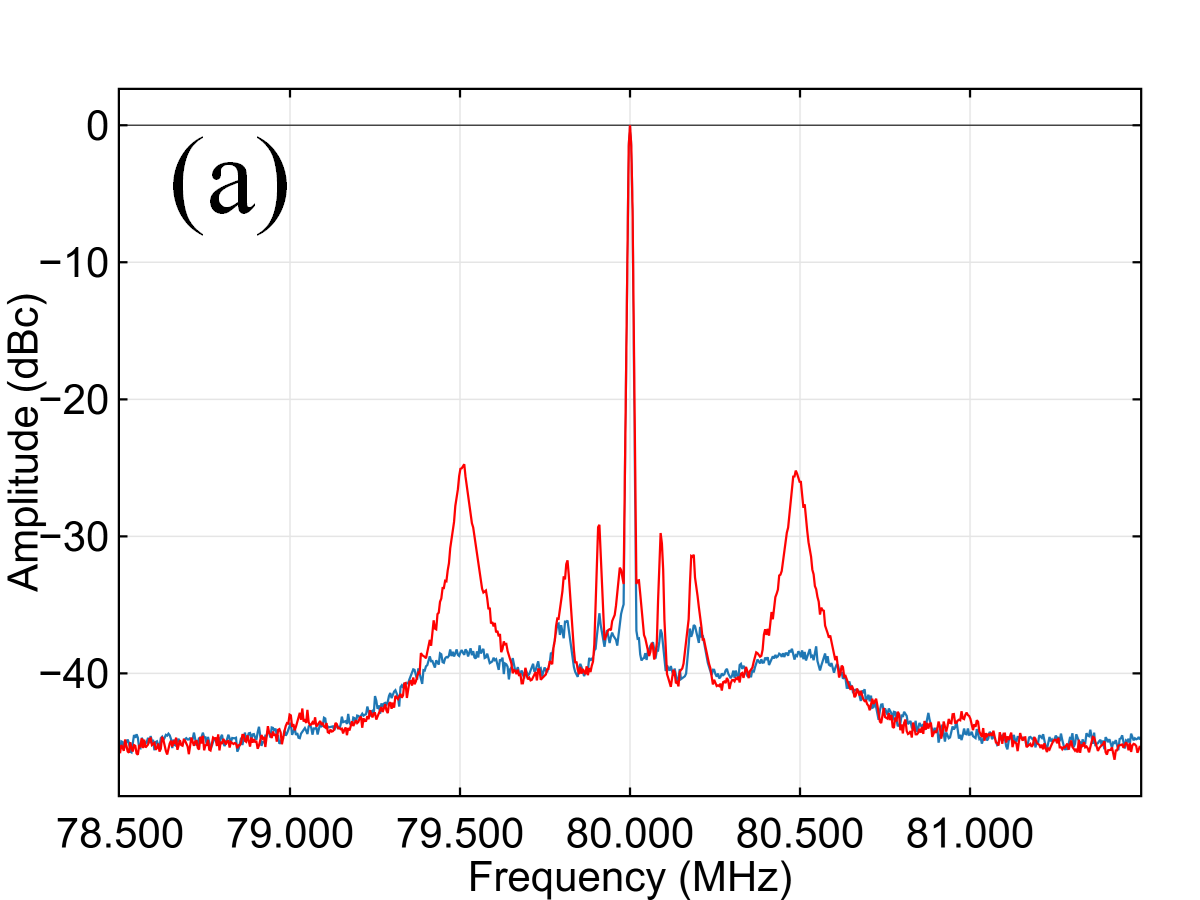}
        \includegraphics[width=0.45\textwidth]{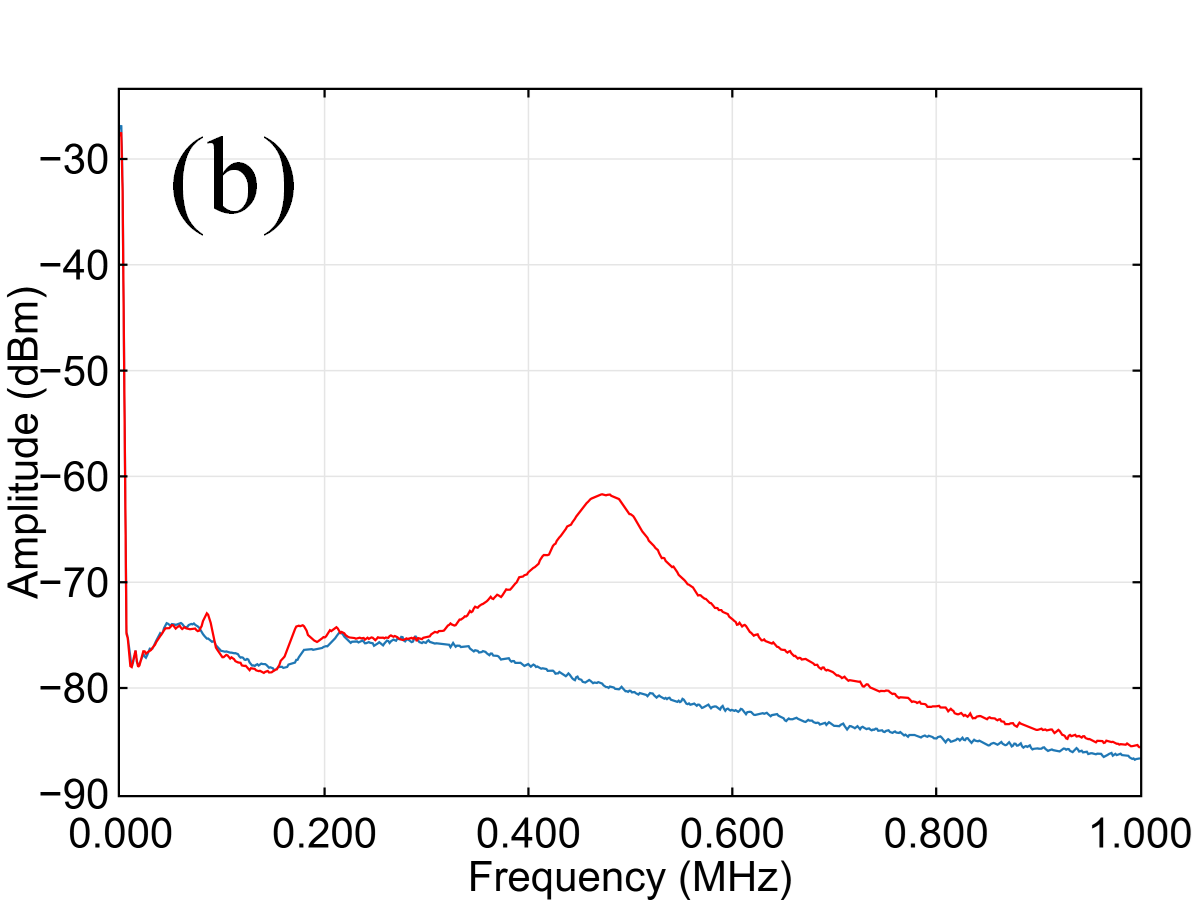}
    \caption{Spectral analysis of the PDH locking system without (red traces) and with (blue traces) the proposed method using Smith predictor. (a) Beat-note signal spectrum centered at the $80\text{ MHz}$ carrier frequency. (b) Error signal spectrum in the locked state.}
    \label{fig:phase noise}
\end{figure}

Based on the estimation using the pure time-delay phase formula:
\begin{equation}
\label{eq:time_delay}
\tau \approx -\frac{\Delta \phi}{\Delta \omega} = -\frac{\phi_2 - \phi_1}{\omega_2 - \omega_1},
\end{equation}
where $\phi_1, \phi_2, \omega_1, \omega_2$ are chosen from Fig. \ref{fig:System Response}(b),
we deduce the delay of the loop outside the FPGA $\tau_\text{out}$ = 106 ns representing the link between the FPGA ports and the laser.
Furthermore, considering that the FPGA-based digital controller employed in the experiment introduces an additional digital computation latency $\tau_2$ during PID algorithm execution, $\tau_2$ must be superimposed onto the total delay for comprehensive compensation when constructing the complete predictive model.
Tailored to the system's frequency response features, the fitting model introduces an inverting leaky integrator in the low-frequency regime, combined with a global first-order low-pass filter structure. 

A set of typical experimental results of implementing the proposed method in the actual PDH frequency stabilization system as Fig. \ref{fig:overall}(b) is presented in Fig. \ref{fig:phase noise}. As shown in Fig. \ref{fig:phase noise}(a), the conventional PDH loop (red trace) exhibits prominent servo bumps at offset frequencies of $\pm 480\text{ kHz}$. We observe that the sharp spikes at $\pm 90\text{ kHz}$ and $\pm 180\text{ kHz}$ are attributed to inherent FPGA noise in our setup. Upon implementing the Smith predictor (blue trace), these servo bumps are effectively suppressed, achieving a $14\text{ dB}$ reduction at the peak positions. Furthermore, the spectrum of the error signal from the locked laser is also measured, Fig. \ref{fig:phase noise}(b) demonstrates that its amplitude at the servo bump frequency is attenuated by $18\text{ dB}$. These results are in agreement with the Radio Frequency (RF) spectrum measurements observed in Fig. \ref{fig:phase noise}(a) and indicate the successful construction of the model for the system. Non-zero residual delay remains in practice because the digital pipeline itself takes a finite number of clock cycles, and its effect on the achievable suppression is quantified in the Supplemental Material.

\begin{figure}[h]
    \centering
    \includegraphics[width=\columnwidth]{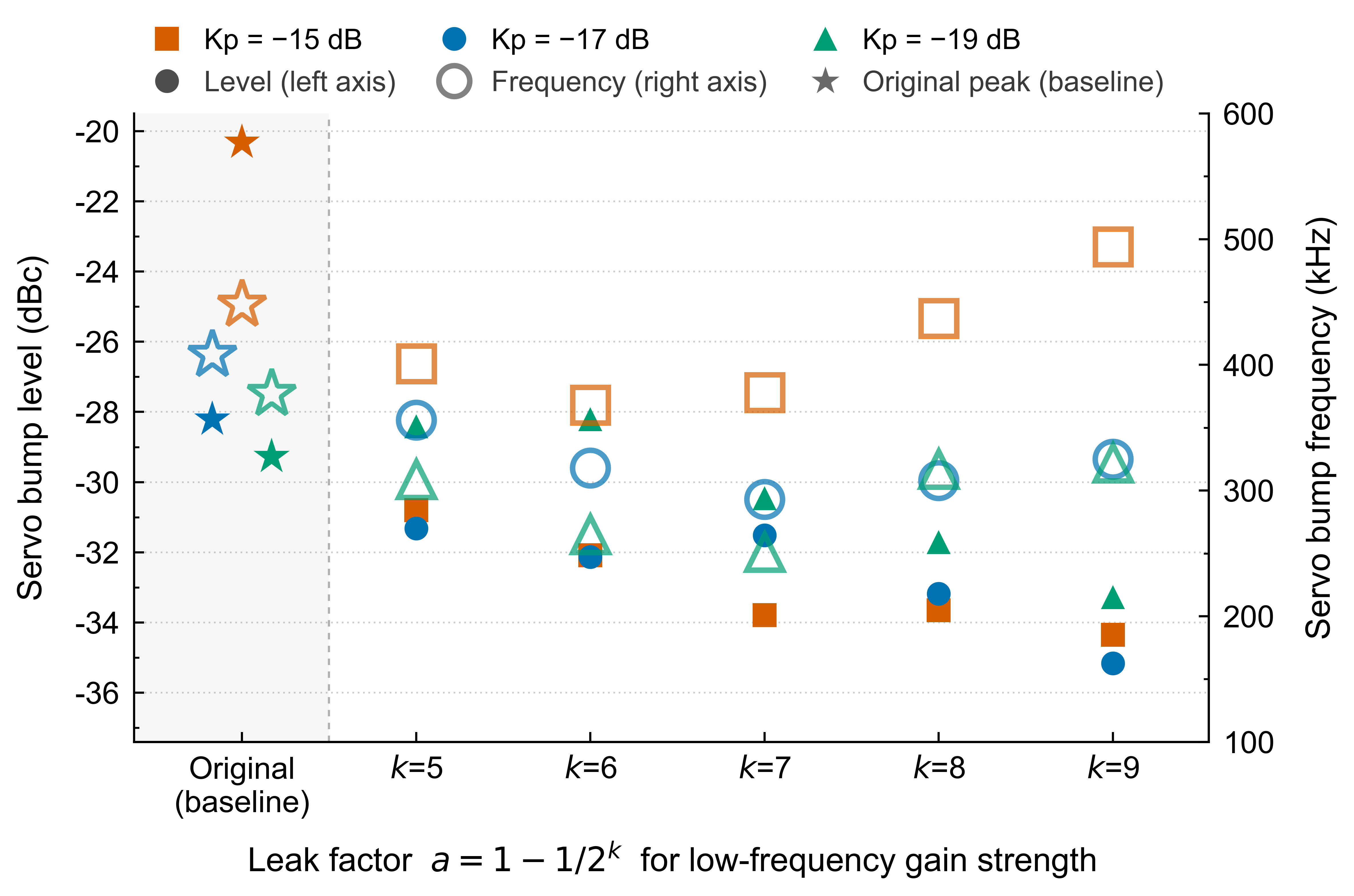}
    \caption{Experimental performance versus key parameters. The left vertical axis denotes the servo-bump level and the right vertical axis its peak frequency. Filled and open markers correspond to the left and right vertical axes, respectively; stars denote the original peak positions.}
    \label{fig:performance_exam}
\end{figure}

We further investigate how the performance of Smith predictor depends on key parameters of modeling and implementation. The locking gain and the degree of deviations between the predictive model and the actual system property affect the suppression of servo bumps. Specifically, the leakage strength of the first-order inverting leaky-integrator stage in the prediction model is set to $k = 5, 6, 7, 8, 9$, where a larger value of $k$ yields a higher low-frequency gain and hence a better match. We have performed measurements at three locking gains, $K_p = -15, -17, -19$ dB, and the results are shown in Fig. \ref{fig:performance_exam}. 
The leftmost column collects the original status of servo bump peaks including -20.33 dB @ 448 kHz, -28.22 dB @ 408 kHz and -29.28 dB @ 377 kHz when the laser system works under three slightly different conditions. The level of all three groups decreases as $k$ increases from small values, reaching -34.35 dB, -35.16 dB and -33.28 dB at $k = 9$. At small values of $k$ the Kp = -19 dB group exhibits the weakest suppression, with a level comparable to that of its original status. Meanwhile, the frequencies of servo bump peaks generally shift positions with changing values of $k$.

\begin{figure}[h]
    \centering
        \includegraphics[width=0.45\textwidth]{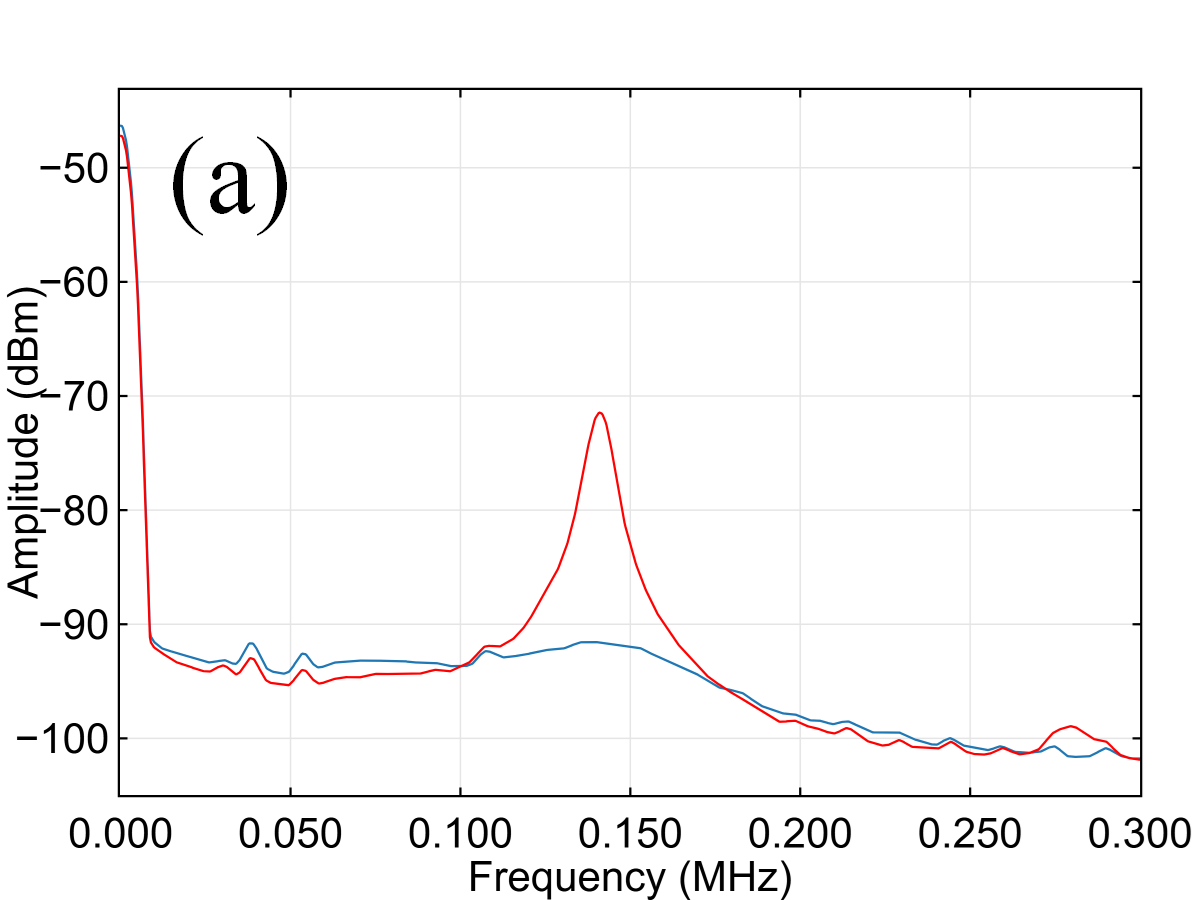}
        \includegraphics[width=0.45\textwidth]{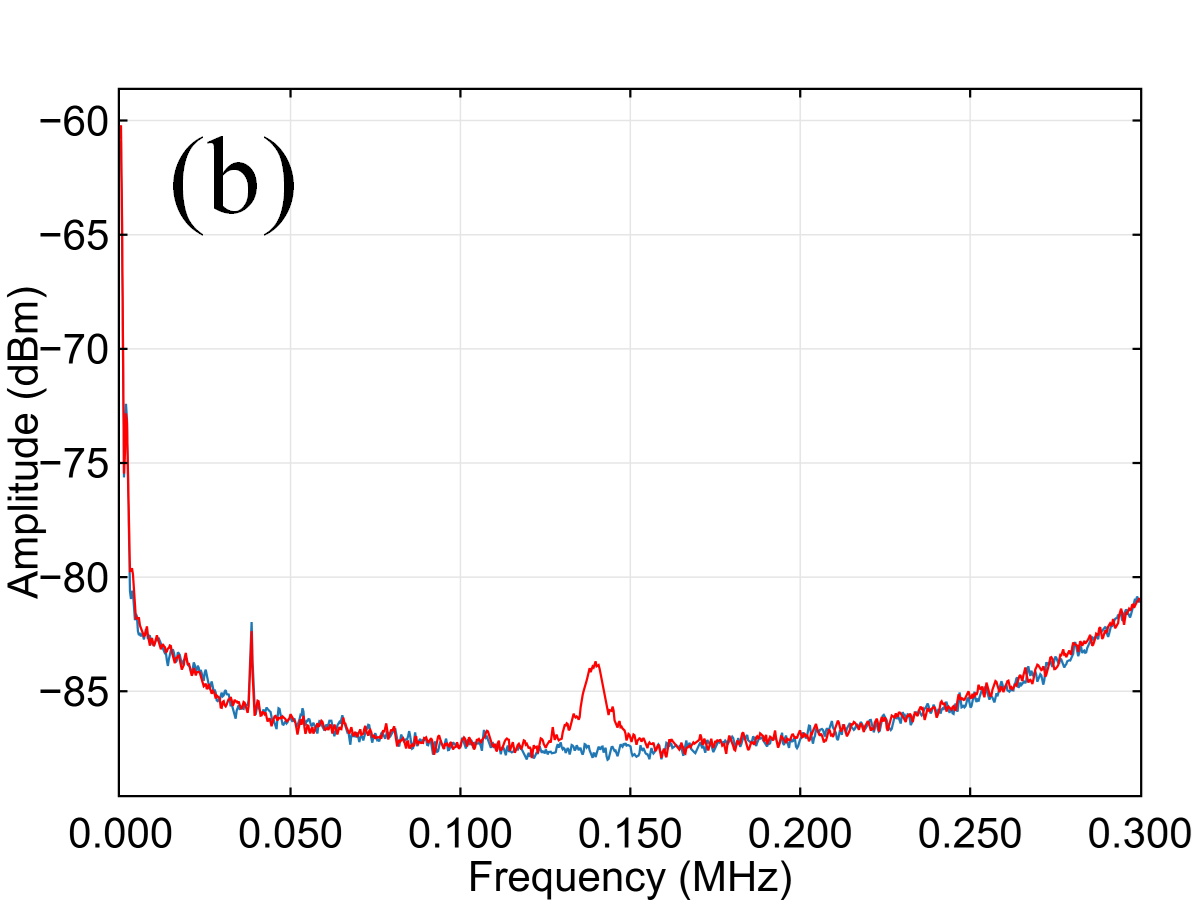}
    \caption{Experimental results of the Rydberg atom quantum sensing. The red and blue traces correspond to the system operating without and with the proposed method of predictive electronic feedback control, respectively. (a) Spectra of error signals of the probe beam in the locked state. (b) Spectra of the transmission signal of probe laser via Rydberg EIT.}
    \label{fig:longwave_sensing}
\end{figure}

As an example of applications with the proposed method of predictive electronic feedback control for narrow-linewidth lasers, we carry out a quantum sensing experiment of low-frequency electric field with Rydberg Electromagnetically Induced Transparency (EIT) \cite{Jin2025CPL, Yuan_polarizatioin_wm} where the probe laser is stabilized with this method. Fig. \ref{fig:longwave_sensing} compares the signals of the system with conventional PDH locking and the proposed method. As can be observed from the red traces in Fig. \ref{fig:longwave_sensing}(a) and \ref{fig:longwave_sensing}(b), the conventional method exhibits a prominent noise bump near 140 kHz in both the baseband error signal and the Rydberg EIT transmission signal of probe laser. Furthermore, a profile analysis reveals a convincingly positive correlation (Pearson $r = 0.9023$) between the spectral envelopes of the error signal and the Rydberg EIT signal, according to discrete amplitude data points of both spectra within the 110 kHz to 160 kHz frequency band. 
This co-variation is anticipated, since the noise of Rydberg EIT process contains contribution from the frequency noise of probe laser. The simultaneous suppression in both channels when the Smith predictor is enabled also provides the evidence that the reduced noise peak in Rydberg EIT signal originates in the servo bump.
Specifically, the initial noise bump of 3.7 dB in the Rydberg EIT signal is eliminated and reduced to the noise floor level. Consequently, if the frequency of the target electric signal is located near the 140 kHz servo bump, employing the proposed method can enhance the SNR of detection by 2.46 dB. Further details are provided in the Supplemental Material.


In conclusion, we have demonstrated that the servo bump issue imposed by finite loop delay in PDH locking can be effectively alleviated within the electronic
feedback path alone. By embedding a Smith predictor in a single FPGA
controller, the model-based compensation reconstructs a virtually delay-free error signal and relaxes the bandwidth limit set by the total loop delay, without
adding optical components or sacrificing available output laser power. Measured
against an 80~MHz optical beat note, the predictor reduces the servo bumps by
14~dB and the locked-state error signal by 18~dB. Deployed in a Rydberg-based
electric-field sensor, it suppresses a 3.7~dB bump near 140~kHz down to the
detector noise floor, improving the signal-to-noise ratio by 2.46~dB. Requiring
no modification to the plant, the method transfers directly to optical clocks,
coherent spectroscopy, and other delay-limited narrow-linewidth platforms.

The authors gratefully acknowledge funding supports from the Strategic Priority Research Program of the Chinese Academy of Sciences (Grant No. XDB1690300), the National Key R\&D Program of China (Grant No. 2024YFB4504002), and the Space Application System of China Manned Space Program.

\bibliographystyle{apsrev4-2}

\bibliography{glacier_ref}
\end{document}